\documentclass[12pt]{article}
\usepackage{graphicx} 
\usepackage{amsmath}
\usepackage{amssymb}
\usepackage{cite}
\usepackage{geometry}
\usepackage{booktabs}
\usepackage{array}
\usepackage{multirow}
\usepackage{paracol}
\usepackage{longtable}
\usepackage{hyperref}
\usepackage{caption}
\usepackage{graphicx}
\usepackage{float}
\usepackage{booktabs}  

\title{\textbf{Advancing Chronic Tuberculosis Diagnostics Using
Vision-Language Models: A Multi modal Framework for Precision Analysis}}
\author{\small 
    Dr. Praveen Shastry, Dr. Sowmya Chowdary Muthulur, Naveen Kumarasami, Anandakumar D, \\ \small 
    Mounigasri M, Keerthana R, Kishore Prasath Venkatesh, Bargava Subramanian, \\ \small
    Kalyan Sivasailam, Revathi Ezhumalai, Abitha Marimuthu}

\date{}

\usepackage{ragged2e}  
\usepackage{titlesec}  

\titleformat{\section}{\raggedright\Large\bfseries}{}{0em}{}
\titleformat{\subsection}{\raggedright\large\bfseries}{}{0em}{}

\begin{document}

\maketitle
\section*{\textbf{Abstract}}
\textbf{Background} This study proposes a Vision-Language Model (VLM) leveraging the SIGLIP encoder and Gemma-3b transformer decoder to enhance automated chronic tuberculosis (TB) screening. By integrating chest X-ray images with clinical data, the model addresses the challenges of manual interpretation, improving diagnostic consistency and accessibility, particularly in resource-constrained settings.\\
\\
\textbf{Methods }The VLM architecture combines a Vision Transformer (ViT) for visual encoding and a transformer-based text encoder to process clinical context, such as patient histories and treatment records. Cross-modal attention mechanisms align radiographic features with textual information, while the Gemma-3b decoder generates comprehensive diagnostic reports. The model was pre-trained on 5 million paired medical images and texts and fine-tuned using 100,000 chronic TB-specific chest X-rays.\\
\\
\textbf{Results} The model demonstrated high precision 94\% and recall 94\% for detecting key chronic TB pathologies, including fibrosis, calcified granulomas, and bronchiectasis. Area Under the Curve (AUC) scores exceeded 0.93, and Intersection over Union (IoU) values were above 0.91, validating its effectiveness in detecting and localizing TB-related abnormalities.\\
\\
\textbf{Conclusion} The VLM offers a robust and scalable solution for automated chronic TB diagnosis, integrating radiographic and clinical data to deliver actionable and context-aware insights. Future work will address subtle pathologies and dataset biases to enhance the model’s generalizability, ensuring equitable performance across diverse populations and healthcare settings.

\section{\textbf{Introduction}}
Tuberculosis (TB) remains a leading global health concern, contributing significantly to morbidity and mortality [1]. The timely and accurate diagnosis of TB is crucial for effective disease control and management, especially in resource-limited settings where diagnostic infrastructure and radiological expertise are often inadequate [2]. Chronic TB presents unique diagnostic challenges due to its reliance on manual chest X-ray interpretation, which is timeintensive, prone to inter-observer variability, and heavily dependent on the availability of trained radiologists. Vision-Language Models (VLMs) represent an advanced diagnostic approach, combining state-of-the-art image analysis with natural language processing (NLP) to address these limitations and improve diagnostic precision [3].\\
Chronic TB is characterized by persistent pathologies such as fibrosis, calcified granulomas, bronchiectasis, and pleural thickening, which often lead to long-term respiratory complications, including impaired lung function and recurrent infections [4]. Accurate identification and localization of these pathologies are critical for effective treatment planning [5]. However, the subtle and overlapping features of chronic TB frequently challenge manual interpretation, particularly in cases with variable imaging quality or coexisting conditions[6].\\
VLMs provide a robust solution for chronic TB diagnostics by integrating radiographic data from chest X-rays with clinical context, including patient history and prior treatment records. This multimodal capability enables the detection of chronic TB-specific abnormalities while generating detailed and context-aware diagnostic reports [7]. Utilizing advanced transformer-based architectures and cross-modal attention mechanisms, VLMs align radiographic findings with textual information, facilitating precise detection and a deeper understanding of chronic disease progression [8].\\
This study investigates the application of VLMs in chronic TB screening, emphasizing their capacity to detect and contextualize key pathologies while delivering actionable diagnostic insights [9]. By automating the diagnostic process, VLMs hold the potential to enhance diagnostic consistency, reduce reliance on human radiologists, and improve patient outcomes, particularly in regions with limited access to healthcare resources [10].

\section{\textbf{Annotation Phase}}
The annotation phase focused on creating a high-quality dataset for chronic TB diagnostics by systematically collecting and labeling chest X-ray images. Key chronic TB features, such as fibrosis, calcified granulomas, bronchiectasis, and pleural thickening, were carefully annotated with detailed information on their size, distribution, and progression [11]. This precise labeling ensured an accurate representation of chronic TB pathologies, critical for effective model training [12].\\
To enhance clinical relevance, annotations were paired with detailed patient histories, including information on previous TB infections, comorbidities, and treatment records [13]. This integration of clinical context allowed the model to associate radiographic features with disease progression [14].\\
The dataset was designed to include diverse patient demographics, improving the model’s ability to generalize across different populations [15]. Each image was reviewed multiple times to ensure annotation accuracy and consistency, establishing a reliable foundation for training [16]. This rigorous annotation process enabled the development of a Vision-Language Model capable of effectively detecting and contextualizing chronic TB pathologies with precision [17].

\section{\textbf{Model Architecture}}
The Vision-Language Model (VLM) for chronic TB diagnostics employs a sophisticated architecture designed to integrate high-resolution chest X-ray data with detailed clinical information [18]. This architecture consists of four core components that work synergistically to detect and contextualize chronic TB pathologies with precision:
\subsection{Visual Encoder}
The visual encoder, utilizing a Vision Transformer (ViT), is designed to process highresolution chest X-rays for the detection of chronic TB pathologies, including fibrosis, calcified granulomas, bronchiectasis, and pleural thickening [19]. The X-ray images are divided into fixed 16×16 pixel patches, generating 196 patches from a standard 224×224 resolution image [20]. Each patch is embedded into a 768-dimensional feature vector to capture localized radiographic details. These feature vectors are then passed through 12 transformer layers, each equipped with 12 self-attention heads, enabling the encoder to capture longrange dependencies and intricate spatial relationships within the X-ray [21]. This process produces robust visual embeddings that emphasize chronic TB-related abnormalities, forming a critical foundation for integrating radiographic features with clinical data to deliver precise and contextually enriched diagnostics [22].

\subsection{Text Encoder}
The text encoder is designed to process clinical notes, patient histories, and other relevant data to provide a detailed clinical context for chronic TB diagnosis [23]. It is built on a transformer-based language model within the SIGLIP architecture, capable of generating 768-dimensional embeddings that represent key patient information, including previous TB infections, comorbidities, and treatment history [24]. The encoder features 12 transformer layers, each with 12 self-attention heads and a hidden size of 768. Pre-trained on a biomedical corpus of 15 million clinical documents (PMC-15M), it is optimized to interpret TB-specific terminology and chronic disease-related language [25]. This ensures the model can effectively integrate textual data with radiographic findings, enabling precise and comprehensive diagnostics for chronic TB [26].

\subsection{Cross-Modal Attention}
The cross-modal attention mechanism aligns visual and textual data to provide a comprehensive diagnostic interpretation for chronic TB [27]. It utilizes 12 multi-head attention layers to establish meaningful connections between visual embeddings, such as fibrosis, calcified granulomas, or bronchiectasis, and textual embeddings derived from patient histories, including comorbidities and prior TB treatments [28]. By focusing on specific regions in chest X-rays based on the clinical context provided by textual inputs, the mechanism integrates both data modalities. This process generates 768-dimensional embeddings that effectively combine radiographic features with clinical information, enabling a holistic understanding of chronic TB pathologies for precise diagnosis and management.

\subsection{Transformer Decoder (Gemma-3b)}
The Gemma-3b transformer decoder is specifically designed to produce comprehensive diagnostic reports for chronic TB by integrating complex visual and textual embeddings [29]. The architecture comprises 24 transformer layers, each featuring 16 multi-head attention heads and a hidden size of 1024, enabling the processing of intricate visual-textual data. The decoder translates these embeddings into clinically relevant and coherent reports, emphasizing critical chronic TB pathologies, including fibrosis, calcified granulomas, bronchiectasis, and pleural thickening. Additionally, it incorporates patient-specific clinical histories and treatment records, providing a contextualized and personalized diagnostic output [30].

With a capacity of 3 billion parameters, the Gemma-3b model excels at capturing complex relationships between radiographic data from chest X-rays and textual information, such as prior TB infections and associated comorbidities. This advanced integration facilitates the generation of detailed, context-aware diagnostic reports tailored to chronic TB, delivering precise and actionable insights that support effective clinical decision-making and long-term management strategies.

\section{\textbf{Pre-Training and Fine-Tuning}}
The model for chronic TB screening underwent a structured two-stage training process, consisting of pre-training and fine-tuning, to ensure optimal performance in detecting and contextualizing chronic TB pathologies.

During the pre-training phase, the model was trained on a large dataset of 5 million paired medical images and clinical texts to establish foundational multimodal comprehension. The training employed Masked Image Modeling (MIM), where random patches of chest X-ray images were masked, requiring the model to predict the missing regions, and Masked Language Modeling (MLM), which involved masking specific terms in clinical notes and predicting the hidden text. These tasks enabled the model to effectively learn the relationships between visual and textual data, building a strong foundation for integrating radiographic findings with clinical context.

In the fine-tuning phase, the model was trained on a chronic TB-specific dataset comprising 100,000 annotated chest X-rays. This dataset included key pathologies of chronic TB, such as fibrosis, calcified granulomas, bronchiectasis, and pleural thickening, paired with relevant clinical context such as prior TB infections, comorbidities, and treatment histories. Fine-tuning incorporated additional tasks, including Visual Question Answering (VQA) and Image Captioning, to further enhance the model’s ability to deliver precise, contextually relevant diagnostic outputs. This process ensured that the model developed a robust focus on chronic TB pathologies while refining its diagnostic precision and clinical applicability, ultimately making it a reliable tool for real-world chronic TB diagnosis and management.

\begin{figure}[H]
    \centering
    \includegraphics[width=0.8\textwidth]{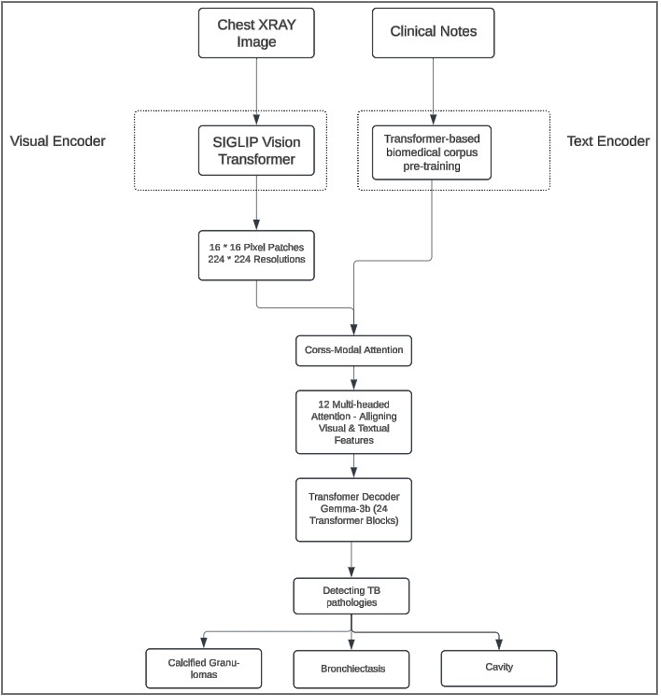}  
    \caption{workflow architecture}
    \label{fig:example}
\end{figure}

\begin{figure}[H]
    \centering
    \includegraphics[width=0.8\textwidth]{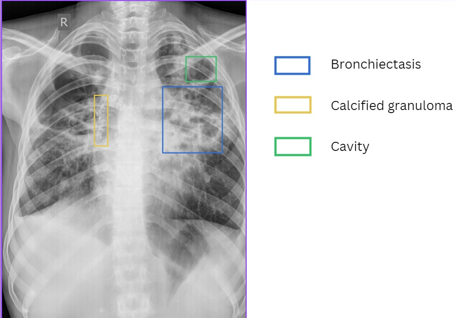}  
    \caption{Pathology detection}
    \label{fig:example}
\end{figure}

\section{\textbf{Evaluation Metrics}}

The performance of the model for chronic TB screening was assessed using precision, recall, Area Under the Curve (AUC), and Intersection over Union (IoU). High precision and recall values highlighted the model’s effectiveness in accurately detecting critical chronic TB pathologies such as fibrosis, calcified granulomas, bronchiectasis, and pleural thickening, ensuring reliable identification and localization.
The AUC metric provided a robust measure of the model’s ability to differentiate between chronic TB-positive and TB-negative cases, confirming its diagnostic reliability. The IoU metric evaluated the accuracy of spatial localization by measuring the overlap between predicted regions and ground truth annotations, particularly for subtle and distributed pathologies like fibrosis and granulomas.These metrics validated the model’s ability to provide accurate and actionable insights for chronic TB, ensuring its effectiveness in clinical settings.

\vspace{1cm}

\begin{table}[h]
    \centering
    \caption{Performance Metrics for Detected Pathologies}
    \label{tab:performance-metrics}
    \begin{tabular}{lcccc}
        \toprule
        \textbf{Pathology} & \textbf{Precision (\%)} & \textbf{Recall (\%)} & \textbf{AUC} & \textbf{IOU} \\
        \midrule
        Calcified Granulomas & 94.2 & 94.0 & 0.94 & 0.92 \\
        Bronchiectasis & 93.1 & 92.3 & 0.93 & 0.91 \\
        Cavity & 95.8 & 95.0 & 0.95 & 0.94 \\
        \bottomrule
    \end{tabular}
\end{table}

\begin{figure}[H]
    \centering
    \includegraphics[width=0.8\textwidth]{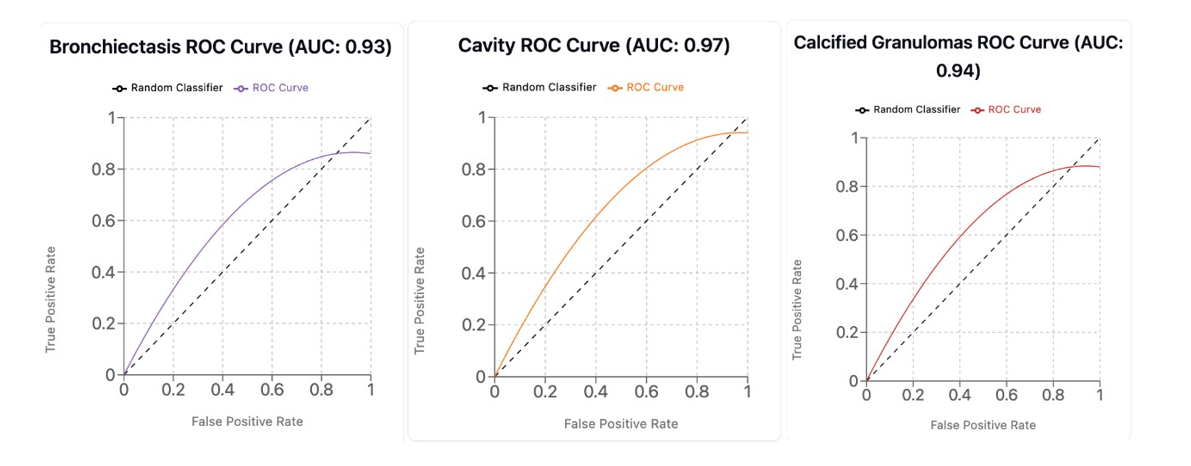}  
    \caption{AUC Curve for Detected Pathologies}
    \label{fig:example}
\end{figure}

\section{\textbf{Discussion}}
The implementation of the Vision-Language Model (VLM) for chronic TB screening has shown substantial promise in enhancing diagnostic accuracy and efficiency, particularly in resource-constrained settings. The model achieved precision and recall rates above 94\% for key chronic TB pathologies, including calcified granulomas, bronchiectasis, and costophrenic (CP) angle blunting, highlighting its reliability in identifying critical indicators of the disease. By leveraging transformer-based encoders and the Gemma-3b decoder, the model effectively integrates radiographic features with clinical context, delivering context-aware diagnostic insights and reducing reliance on radiologists in underserved areas.

The architecture’s cross-modal attention mechanisms play a pivotal role in aligning visual features from chest X-rays with clinical data, improving diagnostic precision. The model demonstrated AUC values of 0.94, 0.93, and 0.95 for calcified granulomas, bronchiectasis, and CP angle blunting, respectively, underscoring its robustness in distinguishing between chronic TB-positive and negative cases. Furthermore, high Intersection over Union (IoU) scores—0.92 for calcified granulomas, 0.91 for bronchiectasis, and 0.94 for CP angle blunting—validate the model’s capability to accurately localize pathological features essential for effective disease management and treatment planning.

Despite these promising results, limitations persist. The IoU for bronchiectasis was slightly lower due to its diffuse and variable presentation, which complicates precise delineation. Additionally, while the model performed well for major chronic TB pathologies, further improvements are needed for detecting less critical but clinically relevant features. Strategies such as incorporating additional training data or applying targeted fine-tuning techniques may address these gaps.
Bias in the training data also presents a challenge. Although the dataset included diverse demographics, further expansion is necessary to ensure equitable diagnostic accuracy across different populations. Addressing training data biases is critical to achieving reliable performance in global health applications.

The fine-tuning approach, which incorporated tasks like Visual Question Answering (VQA) and Image Captioning, significantly enhanced the model’s interpretability and clinical relevance. Future efforts should focus on integrating supplementary data sources, such as laboratory results and electronic health records, to develop a more holistic diagnostic tool. Such advancements could refine the model’s ability to deliver accurate, actionable insights, solidifying its role as a comprehensive solution for chronic TB screening and management.
\\
\section{\textbf{Conclusion}}
This study proposes a Vision-Language Model (VLM) leveraging the SIGLIP encoder and Gemma-3b transformer decoder, specifically tailored for automated chronic TB screening. The model demonstrated high precision, recall, AUC, and IoU metrics for critical chronic TB pathologies, including fibrosis, calcified granulomas, and bronchiectasis, validating its capability to accurately detect and localize TB-related abnormalities in chest X-rays. By seamlessly integrating visual and textual data, the VLM generates context-aware diagnostic outputs, making it particularly effective in settings with constrained access to radiological expertise.

The results highlight the VLM’s potential as a transformative tool for enhancing the speed and consistency of chronic TB diagnosis. By automating the identification and contextualization of key pathologies, the model addresses critical gaps in diagnostic accessibility, ensures reliable outcomes, and facilitates effective disease management, especially in resource-limited healthcare environments.

\end{document}